1# In commemoration of the 85$^{th}$ birthday of Edwald Abramovitch Zavadskii
## (1927-2005)

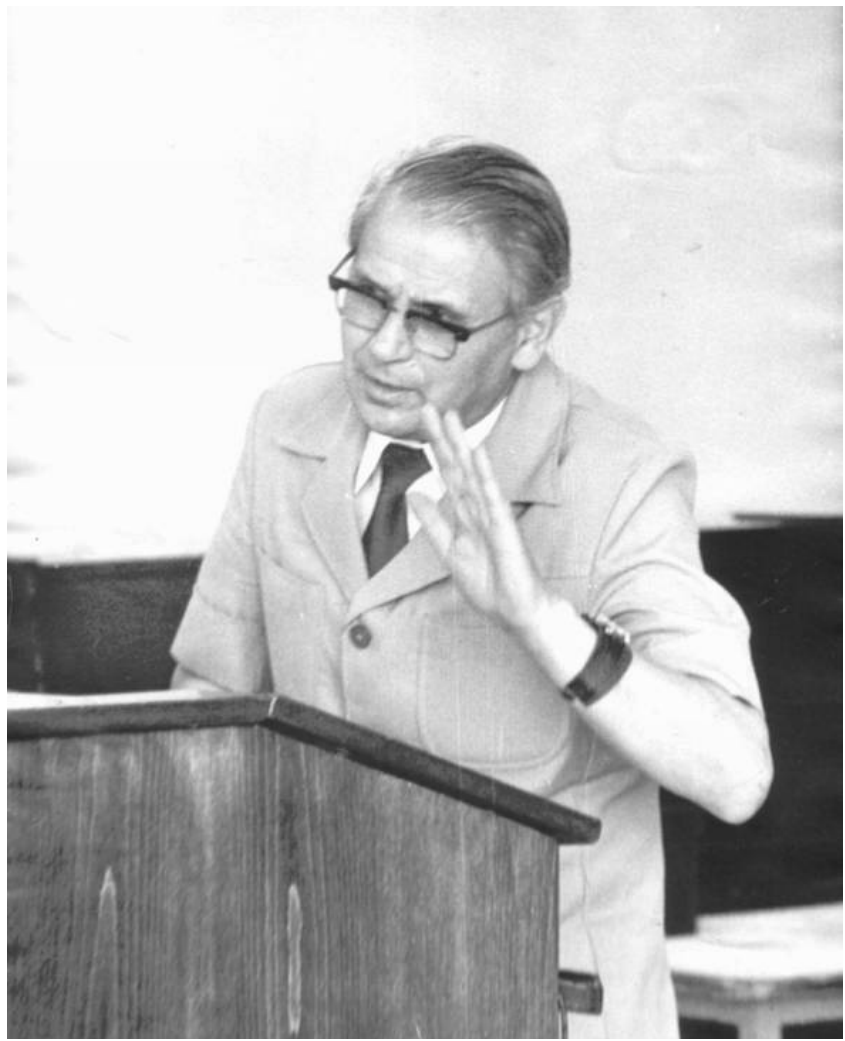

On June 2, 2012 it would have been the eighty fifth birthday of Edwald Abramovitch Zavadskii (1927 –2005), corresponding member of the National Academy of Sciences, brilliant experimental physicist and a person with a very uncommon and vibrant life..

A famous scientist, a talented researcher, a real physicist, a founder of the school of physics of magnetic phenomena at Donetsk Institute for Physics and Engineering, an outstanding teacher and person, Edwald Abramovich Zavadskii was born in 1927 in a German settlement of Millerovo (Rostov Region, Russia) into a family of technical intellectuals.

When the Great Patriotic War began in 1941 and the Germans of the settlement became the "enemies of the system", the normal life of Edwald, 14-years old teenager at the time, ended. His family was exiled to Orsk, Orenburg Region, where they lived for 14 years "behind the barbed wire". Edwald Abramovich did not like to talk about those years: "It is beyond one's nerves". The only relief which could be found in these conditions would come from the many intelligent, highly educated people who were among those 100 thousand exiles, so they taught how to live by their own example and helped those who wanted to study and to learn. And so, Edwald as a distance learner graduated first from the school, then the technical secondary school and finally the Physics Department of Orsk Educational Institute.

Only in 1955 the restrictions on movement for E.A. were lifted and he enrolled for graduate studies at the Institute for Physics of Metals, Academy of Sciences of USSR, in Sverdlovsk. In 1961, he defended his thesis "Galvanomagnetic properties of germanium in high pulsed magnetic fields" supervised by Prof. I.G. Fakidov.

In 1966, after having been invited by Academician A.A. Galkin, E.A. Zavadskii lead the research in magnetic phenomena at Donetsk Institute for Physics and Engineering, NAS of Ukraine. Until the very last days of his life, Edwald Abramovitch associated his creative work with this institute.

Here he started from the position of the head of department and grew up to the position of the Director of the Institute. In 1978, he became a corresponding member of NAS of Ukraine. He founded an academic scientific journal, High Pressure Physics and Technics and since 1987 he was its Editor-in-Chief.

A devotee of science. The international recognition came to E.A. Zavadskii through a new approach to the analysis of magnetic and structural properties of materials at high pressures and in high magnetic fields. He first introduced the concept of metastable phases hidden in the area of negative pressures and developed methods to induce them by applying high magnetic field in ferromagnets.

Another new trend was the use of pulsed magnetic field spectroscopy for establishing the distribution of higher terms of 3-d ions on the basis of magnetic tests of insulators with strong single-ion anisotropy in high pulsed magnetic fields.

E.A. Zavadskii was awarded orders and medals for his outstanding educational and scientific work. He was awarded the K.D. Sinelnikov Academic Prize and the "Order of the Red Banner of Labor". E. A. Zavadskii was an author of more than 200 publications, 4 books. His acted as a supervisor for 21 candidates and 4 doctors of sciences.

Until his last days, he was at the scientific post. He worked as the Director`s Advisor, took an active part in the formation of the overall research programme of the Institute, led the research at the Department of the magnetic properties of solids.

We will always remember his intelligence, kindness, sensitivity, ability to work and help people. He generated creative atmosphere, igniting people and uniting them in the scientific quest.

**The main areas of research interests of Prof. E.A. Zavadskii.**

1. Development of a method for generating record high pulsed magnetic fields (up to 750 kOe); detection of irreversible martensitic transformation in austenitic steels, induced by pulsed magnetic field exceeding a certain threshold value. The latter result had acquired a particular significance after the discovery of induction of states hidden in the area of negative pressures.

These results were awarded the gold medal of the All-Union Exhibition of Achievements of National Economy in 1961.

2. Experimental proof of task-oriented change of the type of magnetic ordering in the sample exposed to high pressure and classification of the possible types of P-T diagrams of magnetically ordered materials.

3. Detection and explanation of the nature of wide metastable areas in temperature – magnetic field – pressure phase diagrams in a number of alloys and compounds of transition metals (the temperature range of 100-200 K and the pressure range of 10 kbar) and anomalously large shifts of the limits of these areas in magnetic field (100 kOe-10 kbar).

4. Discovery and study of the effect of induction of the magnetic phases hidden in the area of negative pressures. The exposure of a material to strong magnetic field can induce and stabilize the magnetic phases that would occur spontaneously without field only in the region of negative pressures.

E. A.'s lead in the discovery of this effect was acknowledged by a number of other laboratories.

When studying the nature of the stability of in the induced phase, the crucial role of nucleation and heterogeneous states was established. This achievement became possible due to successful co-operation of experimentalists of Donetsk Institute for Physics and Engineering and NAS Academician V.G. Bar'yakhtar and his disciples.

5. Development of a method for determining the boundaries of magnetic phases at negative pressures.

The scientific results obtained by Prof. E.A. Zavadskii were awarded the second gold medal of the All-Union Exhibition of Achievements of National Economy and the K.D. Sinelnikov Academic Prize. The most complete description of this work is presented in the following monographs:

E.V. Kuzmin, G.A. Petrakovskii, E.A. Zavadslii. Physics of magnetically ordered materials. - Moscow: Nauka, 1976, 279 pages.

E.A. Zavadslii, V.I. Valkov. Magnetic phase transitions. - Naukova Dumka, 1980, 196 pages.

6. The idea of inducing hidden magnetic states by magnetic field was extended to ferroelectric materials, in which the electric field can bring about a ferroelectric zone, which cannot be realized in these materials otherwise.

The research formed the main part of the book:

E.A. Zavadslii, V.M. Ischuk. Metastable states in ferroelectrics. - Naukova Dumka, 1987, 256 pages.

7. Experimental detection of wide metastable areas in transition metal fluorosilicates and the proof that it is possible to form the prescribed type of magnetic ordering (ferromagnetism or antiferromagnetism, if desired) in a given region of the phase diagram by following a particular sequence of pressure and temperature changes.

8. Fruitful cooperation with Prof. Klaus Bärner, Georg - August University, Göttingen, resulted in the book:

K. Bärner, E.A. Zavadskii Induzierte magnetische Zustände. Shaker Verlag. Aachen, 2007, 306 Seiten.

The above results demonstrate only some of the milestones in the research activities aimed at creation of materials with desired properties that were successfully developed by the Corresponding Member of NASU E. A. Zavadskii and his students.

Alongside his scientific activities, E. A. Zavadskii was doing a great organizational work both as a member of several scientific, qualifying and coordinating councils, and as a university professor. He dedicated a lot of his attention to the strengthening of ties between Donetsk Institute for Physics and Engineering of the National Academy of Sciences of Ukraine and the Donetsk State University: a branch of the University department functioned successfully at the Institute. The graduates of the University provide a reliable supply of new staff to the research departments. For 9 years, E. A. Zavadskii was the deputy director for research, and then he headed the institute for 10 years.

The authors are very grateful to Profs. A.V. Leont'eva, V.I. Valkov, V.I. Kamenev for the discussion and valuable advice.

V.V. Eremenko, Academician of NAS of Ukraine, Kharkov, B.Verkin Institute for Low Temperature Physics and Engineering of the NAS of Ukraine, Ukraine

V.N. Varyukhin, Corresponding Member of NAS of Ukraine, Donetsk, Donetsk Institute for Physics and Engineering named after A.A. Galkin, NAS of Ukraine, Ukraine

V.F. Klepikov, Corresponding Member of NAS of Ukraine, Kharkov, Institute of Electrophysics & Radiation Technologies of the NAS of Ukraine, Ukraine

Prof.V. Beloshenko, Donetsk, Donetsk Institute for Physics and Engineering named after A.A. Galkin, NAS of Ukraine, Ukraine

Prof. E. Feldman, Donetsk, Institute for Physics of Mining Processes, NAS of Ukraine, Ukraine

Prof. A. Leont'eva, Haifa, Israel

Prof. Yu. Mamaluy, Donetsk, State University, Ukraine

Prof. G. Levchenko, Donetsk, Donetsk Institute for Physics and Engineering named after A.A. Galkin, NAS of Ukraine, Ukraine

Prof. Yu. Medvedev, Donetsk, Donetsk Institute for Physics and Engineering named after A.A. Galkin, NAS of Ukraine, Ukraine

Prof. A. Prokhorov, Donetsk, Donetsk Institute for Physics and Engineering named after A.A. Galkin, NAS of Ukraine, Ukraine

Prof. V. Yurchenko, Donetsk, Donetsk Institute for Physics and Engineering named after A.A. Galkin, NAS of Ukraine, Ukraine

Prof. V. Val`kov, Donetsk, Donetsk Institute for Physics and Engineering named after A.A. Galkin, NAS of Ukraine, Ukraine

V. Kamenev, Donetsk, Donetsk Institute for Physics and Engineering named after A.A. Galkin, NAS of Ukraine, Ukraine

V. Ishchuk, Donetsk, Donetsk Institute for Physics and Engineering named after A.A. Galkin, NAS of Ukraine, Ukraine

Prof. Yu. Zavorotnev, Donetsk, Donetsk Institute for Physics and Engineering named after A.A. Galkin, NAS of Ukraine, Ukraine

B. Todris, Donetsk, Donetsk Institute for Physics and Engineering named after A.A. Galkin, NAS of Ukraine, Ukraine

A. Sivachenko, Donetsk, Donetsk Institute for Physics and Engineering named after A.A. Galkin, NAS of Ukraine, Ukraine

# List of the main publications of Prof. E.A. Zavadskii

## Monographs

1. E.V. Kuzmin, G.A. Petrakovskii, E.A. Zavadslii. Physics of magnetically ordered materials. - Moscow: Nauka, 1976, 279 pages.
2. Zavadslii E.A., Valkov V.I. (1980). Magnetic phase transitions. Naukova Dumka. Kiev. 196 pages.
3. Zavadslii E.A., Ischuk V.M. (1987). Metastable states in ferroelectrics. Naukova Dumka. Kiev 256 pages.
4. Bärner K., Zavadskii E. A. (2007). Induzierte magnetische Zustände. Shaker Verlag. Aachen. 306 pages.

## Papers

1. Fakidov I.G., Zavadskii E.A. (1958). The oscillation of the electric resistance n-type germanium in high pulsed magnetic fields. *Soviet Physics JETP*, 7, 716.
2. Fakidov, I.G.; Zavadskii, E.A. (1959). Generator of super-strong pulsed magnetic fields. *Fiz. Metal Metalloved.*, 8, 562.
3. Zavadskii E.A., Fakidov I.G. (1960). Electrical conductivity of n-Ge in a high magnetic field. *Fiz. Metal Metalloved.*, 10, 180.
4. Zavadskii E.A., Fakidov I.G. (1961). Change in electrical conductivity n-Ge in strong pulsed magnetic fields. *Fiz. Metal Metalloved.*, 11, 141.
5. Zavadskii E.A., Fakidov I.G. (1961). Magnetization of the intermetalic compound $MnAu_2$ in high magnetic fields. *Fiz. Metal Metalloved.*, 12, 832.
6. Zavadskii E.A., Fakidov I.G., Samarin N.Y. (1965). Magnetic susceptibility of antiferromagnetics $MnSO_4$, MnO and FeO in high magnetic fields. *Soviet Physics JETP*, 20, 558.
7. Galkin A.A., Zavadskii E.A., Morozov E.M. (1968). New magnetic transformation in the system $Mn_2Fe_ySb_{1-y}$. *JETP Letters*, 8(11), 400-402.
8. Galkin A. A., Zavadskii E. A., Morozov E. M. (1970). Magnetic transformation in the $Mn_2Ge_ySb_{1-y}$ system in strong magnetic fields under high pressure. *Physica Status Solidi (B), 37*(2), 851-856.
9. Galkin A. A., Zavadskii E. A., Val`kov V.I. (1971). Phase transformation in MnAs induced by a strong magnetic field. *Physica Status Solidi (B), 46*(1), K23-K25.
10. Zavadskii E. A., Medvedeva L. I. (1974). Magnetic transitions on the manganese–rhodium system. *Sov Phys Solid State, 15*(8), 1595-1598.
11. Galkin A.A., Zavadskii E.A., Smirnov V.M., Val'kov V.I. (1974). Observation of ferromagnetic state in antiferromagnetic alloys of the $Mn_{1-x}Fe_xAs$ system. *JETP Letters* 20(4), 111-112.